\documentclass[twocolumn,pre,showpacs,preprintnumbers,amsmath,amssymb]{revtex4}

\usepackage{graphicx}
\usepackage{dcolumn}
\usepackage{bm}

\begin{document}
\title{\textbf{\textrm{Time parameterization and stationary distributions in a relativistic
gas}}}

\author{Malihe Ghodrat}

\author{Afshin Montakhab}
    \email{montakhab@shirazu.ac.ir}

\affiliation{Department of Physics, College of Sciences, Shiraz
    University, Shiraz 71454, Iran}

\date{\today}
\begin{abstract}
In this paper we consider the effect of different time
parameterizations on the  stationary velocity distribution function
for a relativistic gas. We clarify the distinction between two
such distributions, namely the J\"{u}ttner and the modified
J\"{u}ttner distributions. Using a recently proposed model of a
relativistic gas, we show that the obtained results for the
proper-time averaging does not lead to modified J\"{u}ttner
distribution (as recently conjectured), but introduces only a Lorentz factor $\gamma$ to the
well-known J\"{u}ttner function which results from observer-time
averaging. We obtain results for rest frame as well as moving
frame in order to support our claim.
\end{abstract}

\maketitle
\textit{Introduction--}Following a maximum entropy principle,
Ferencz J\"{u}ttner \cite{Jut11} presented the first relativistic
generalization of Maxwell-Boltzmann (MB) distribution in 1911.
Although J\"{u}ttner distribution has been generally accepted and
used in high energy and astrophysics
\cite{Ter71,Sun72,Lib90,Ito98,Cer02,Die06,Hee06}, some authors
suggested several alternatives \cite{Hor81,Hor89,Sch05,Leh06} that
can be summarized in terms of the following $\eta$-parameterized
probability distribution functions (PDFs),

\begin{equation}\label{Eq.0}
    f_{\eta}(\mathbf{p})=\frac{\exp(-\beta E)}{\mathcal{Z}E^{\eta}},
\end{equation}
in which $E=(m^{2}+\mathbf{p}^{2})^{1/2}=m\gamma(v)$ is the
relativistic single-particle energy with $c=1$, rest
mass $m$ and Lorentz factor $\gamma(v)=(1-v^{2})^{1/2}$.
$\mathcal{Z}$ is the normalization constant and $\beta$ is the
temperature parameter. For $\eta=0$ and $\eta=1$, the above PDFs
reduce to the J\"{u}ttner function and the so called modified
J\"{u}ttner function, respectively.

\smallskip
The lack of rigorous microscopic derivations and experimental
evidences, made it difficult to decide which of the proposed
relativistic distributions is the correct generalization of MB
distribution. To resolve the uncertainty, semi-relativistic
\cite{Ali06} and fully relativistic \cite{Cub07,Mon09} molecular
dynamics simulations as well as Monte Carlo studies \cite{Pea09}
have been performed by different groups in recent years that
unequivocally favored J\"{u}ttner distribution. However, some
recent investigations on relativistic Brownian motions
\cite{Dun09,Han09} have revealed that stationary distributions can
differ depending on the underlying time-parameterizations, a
problem which never arises in Newtonian physics due to the
existence of a universal time for all inertial observers. On the
other hand, the maximum relative entropy principle (MREP)
\cite{Och76,Ochs76,Weh78} depicts how symmetry considerations lead
to different stationary distributions, each with its own merits
\cite{Dun07}. Arguments of this kind suggest that one may possibly
seek multiple relativistic generalizations of MB distribution.
Using molecular dynamics simulations \cite{Cub09} and theoretical
analysis \cite{Dun09}, some authors have recently proposed that it
is possible to establish a connection between the time-parameter
and the kind of symmetry that lead to a special distribution.
Consequently, relativistic distributions fall into two classes
that are distinguished by their associated symmetry and
time-parameter. We, however, believe that the entire issue calls
for a more careful reconsideration.

\smallskip
In this article, we will first review how the choice of time
parameters or reference measures affect the resulting stationary
distribution. We further clarify the resulting consequence of these different choices.
Using our recently proposed two dimensional hard
disk model \cite{Mon09}, we perform simultaneous measurements at
fixed observer's time `$t$' and particle's proper time `$\tau$'
with respect to a rest as well as a moving observer. The obtained
results are then compared with stationary distributions consistent
with different symmetry considerations. $t$-parametrization is shown to lead to
J\"{u}ttner distribution, as is well-known. However, we find that
$\tau$-parametrization, leads to a PDF that is consistent with a
J\"{u}ttner function divided by a $\gamma$ term, which is
decidedly different from  the original modified J\"{u}ttner
distribution obtained from MREP. This distinction comes more to light when one considers a moving frame.

\smallskip
\textit{Stationary distributions and symmetry considerations--}
The idea of maximizing relative entropy with respect to a
pre-specified measure was a major step forward to clarify the
underlying mathematical differences of the two mostly cited
generalizations of MB distribution, i.e., J\"{u}ttner and modified
J\"{u}ttner functions \cite{Dun07}.

\smallskip
Relative entropy \cite{Weh78} characterizes probability
distributions with respect to a specific reference density,

\begin{equation}\label{Eq.1}
S[f|\rho]\equiv -\int d^{d}\mathbf{p} f(\mathbf{p}) \log
\frac{f(\mathbf{p})}{\rho(\mathbf{p})},
\end{equation}
in which, $\rho(\mathbf{p})>0$ plays the role of a reference
density. It is then possible \cite{Dun07} to develop a general,
coordinate invariant form of maximum entropy principle, under the
constraints

\begin{eqnarray}
1&=&\int d^{d}\mathbf{p} f(\mathbf{p})\\ \label{normconst}
\epsilon=E_{tot}/N&=&\int d^{d}\mathbf{p}\label{eneconst}
f(\mathbf{p})E(\mathbf{p}).
\end{eqnarray}
The resulting stationary distribution is then given by,
\begin{equation}\label{Eq.3}
   f(\mathbf{p})=\rho(\mathbf{p})\exp(-\beta E(\mathbf{p}))/\mathcal{Z},
\end{equation}

\smallskip
Choosing a constant reference density,
$\rho(\mathbf{p})=\rho_{0}$, the above equation reduces to the
well known J\"{u}ttner function. This choice correspond to a
Lebesgue integrating measure, $d\mu=d^{d}\mathbf{p}$, which is the
only translational invariant measure in the momentum space
\cite{Dun07}. Hence, it implies that J\"{u}ttner distribution is
associated with a reference measure that has translational
symmetry. Since this kind of symmetry is not relevant in
relativistic mechanics, it is interpreted as a momentum
conservation law during collisions.

\smallskip
On the other hand, a momentum dependent reference density,
$\rho(\mathbf{p})=1/E(\mathbf{p})$,  leads to the modified
J\"{u}ttner distribution. The integration measure consistent with
this choice assigns to any subset $\mathcal{A}$ in momentum space,
the measure number
\begin{equation}\label{Eq.4}
    \chi(\mathcal{A})=\int_{\mathcal{A}}d\mu=\int_{\mathcal{A}}\frac{d^{d}\mathbf{p}}{E(\mathbf{p})}
\end{equation}
Considering the fact that $d^{d}\mathbf{p}/E(\mathbf{p})$ is an
invariant quantity under Lorentz transformation, the modified
J\"{u}ttner distribution is associated with a reference measure
that is invariant under the fundamental symmetry group of
relativity.

\smallskip
It should be noted that we have no rigorous theoretical analysis
or experimental evidence to favor one reference density/measure to
the other. Numerical studies are also not decisive due to the
arbitrariness in the choice of time parameters \cite{Cub09,Dun09}
or discretization rules \cite{Dun05,JDun05,JDun07}. Therefore, at this stage we
are obliged to accept the arbitrariness in the choice of reference
density. However, once the reference density is chosen, other
parameters like $\mathcal{Z}$ and $\beta$ are deterministically
obtained using the pre-specified constraints (Eqs. 3 and
\ref{eneconst}). We will show how misleading conclusions can arise
if one fails to appreciate this important point.

\smallskip
\textit{Stationary distributions and time parameters--} Discarding
the classic notion of universal, absolute time introduces
complexity to almost all time-dependent subjects in physics. Among
these, are the evolution of dynamical systems towards equilibrium
and more specifically their equilibrium probability distribution
function (PDF). As reported recently \cite{Dun09,Han09,Cub09}, the
description of the motion of relativistic particles with respect
to the observer's time, $t$, and the particle's proper time,
$\tau$ are completely different. To elucidate, consider the
$t$-averaged PDF of a Brownian particle in terms of the observer's
time
\begin{equation}\label{Eq.5}
f_{t}(v)=\frac{1}{t}\int_{0}^{t}dt'\delta(v-\mathbf{V}(t')).
\end{equation}
The velocity of the particle, $\mathbf{V}(t)$, can also be
parameterized in terms of the particle's proper time denoted by
$\hat{\mathbf{V}}(\tau)$. The corresponding $\tau$-averaged PDF is
then defined as
\begin{equation}\label{Eq.6}
\hat{f}_{\tau}(v)=\frac{1}{\tau}\int_{0}^{\tau}d\tau'\delta(v-\hat{\mathbf{V}}(\tau')).
\end{equation}

By simply changing the integration variable, it can be shown
\cite{Cub09} that the relation between stationary
($t,\tau\rightarrow\infty$) distributions is
\begin{equation}\label{relation}
    \hat{f}_{\infty}(v)\propto\frac{f_{\infty}(v)}{\gamma(v)}
\end{equation}

\smallskip
The appearance of the factor $\gamma(v)$, in the denominator of
Eq.(\ref{relation}) suggests that the $\tau$-averaged distribution
is a modified J\"{u}ttner, if the $t$-averaged distribution turns
out to be J\"{u}ttner function \cite{Dun09,Cub09}. Although the
similarity is very suggestive, in this paper we show that a
J\"{u}ttner function which is simply divided by
$E(\mathbf{p})(=m\gamma(v))$ is not necessarily equivalent to a
modified J\"{u}ttner function.

\smallskip
In order to distinguish the correct interpretation of
Eq.(\ref{relation}), we will check the validity of this equation
by comparing the numerically obtained equilibrium distributions
with respect to different time parameters, in the rest frame.
Next, the results are further discussed by considering the problem
from the viewpoint of a moving observer. In the light of these
numerical simulations, one can decide whether the $\tau$-averaged
distribution is truly described by a modified J\"{u}ttner
function.

\smallskip
The model we have used here is the previously proposed two
dimensional relativistic hard-disk gas \cite{Mon09}. In this
model, the disk-like particles move in straight lines at constant
speed and change their momenta instantaneously when they touch at
distance $\sigma$. The exchange of energy and momenta is governed
by the relativistic energy-momentum conservation laws. In our
simulation we have used $N$ particles of equal rest masses $m$
that are constrained to move in a square box of linear size $L$
with periodic boundary condition. In order to simulate a
stationary system in the rest frame, the center-of-mass momentum
is put to zero manually. This condition would automatically be
satisfied (if not at each instant but at least on average) if
fixed reflecting walls were used \cite{Hai92}. The density is
chosen to correspond to a dilute gas.

\smallskip
In the next two sections we obtain $t$-averaged as well as
$\tau$-averaged PDFs directly from simulations of such a model and
compare them with the various proposed PDFs in order to clarify
the relevance of various PDFs.

\begin{figure}

  \includegraphics[width=8cm]{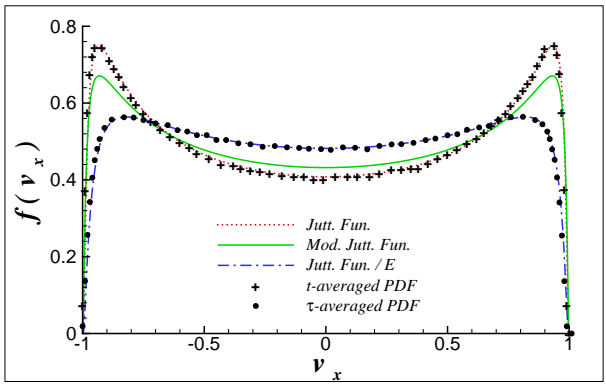}\\
  \caption{Equilibrium velocity distributions in the rest frame:
  numerically obtained $x$ component of $t$-averaged (+) and $\tau$-averaged ($\bullet$) velocity
  distribution from simulation of $N=100$ particles of mass $m=0.1$.
  Here, $\varepsilon=3.06m$ and the corresponding temperature parameters are
  $\beta_{J}=7.62$, $\beta_{MJ}=4.86$. A significant deviation from modified J\"{u}ttner function is
  evident. Similar results are obtained for $f(v_{y})$.}
\end{figure}

\smallskip
\textit{Rest frame--} To obtain the $t$-averaged PDF we let the
system equilibrate (typically after $10^{2}N$ collisions) and
simultaneously measure velocities of all particles at a given
instant of time $t$. To obtain the $\tau$-averaged PDF the
proper-time of each particle, $\tau_{i}\,(i=1,...,N)$, is computed
during the simulation and velocities are recorded at a fixed
proper-time value, $\tau_{i}=\tau$. That is, velocities are
measured when the particles have the same life-time. To collect
more data in the former (latter) scenario, we may either repeat
the procedure for different initial conditions or perform
measurements at several equally-separated time (proper time)
instances.

\smallskip
Simulation results for $N=100$ particles are presented in Fig.1.
Particles are initially placed on a square lattice of constant
$L/\sqrt{N}$ and velocities are chosen randomly and the mean
energy per particle is $\varepsilon=3.06 m$. Velocities are all
measured with respect to a rest observer at $2\times 10^{7}$
instants using time intervals $10^{2}\mathcal{T}$, where
$\mathcal{T}$ is the system's mean free time. As shown in Fig.1,
the $t$-averaged (+) and $\tau$-averaged ($\bullet$) PDFs agrees
well, respectively, with J\"{u}ttner function with $\beta=7.62$
(dotted red line) and J\"{u}ttner function with the same $\beta$,
divided by `$E$' (dashed blue line). This result confirms the
relation described in equation (\ref{relation}). The more
important question, however, is the correct interpretation of the
right hand side of this equation.

\smallskip
Considering the fact that the modified J\"{u}ttner function
differs from J\"{u}ttner function by a factor `$1/E$', one may
conclude that $\tau$-averaged PDFs are described by modified
J\"{u}ttner function in the same manner that $t$-averaged PDFs are
depicted to be J\"{u}ttner function \cite{Cub09}. Before drawing
such a conclusion, we note that by substituting modified
J\"{u}ttner function in Eq.(\ref{eneconst}), one obtains a
relation between the temperature parameter, $\beta$, and the conserved
quantities of the system ($\epsilon,m$), which is different from
the one obtained if one substitutes J\"{u}ttner function instead.
For the two dimensional momentum space we have:
\begin{eqnarray}
  \epsilon&=&\frac{2}{\beta_J}+\frac{m^{2}\beta_{J}}{1+m\beta_{J}}\\
  \epsilon&=&m+\frac{1}{\beta_{MJ}},\label{beta}
\end{eqnarray}
where the indexes $J$ and $MJ$ refers to J\"{u}ttner and modified
J\"{u}ttner, respectively. In our simulation, this leads to
$\beta_{J}=7.62$ and $\beta_{MJ}=4.86$. Therefore, simply dividing
a J\"{u}ttner function by $\gamma$ does not lead to a modified
J\"{u}ttner function. The functional form as well as the
temperature parameter are important in distinguishing a modified
J\"{u}ttner function from its counterpart. This point is clearly
demonstrated in Fig.1 as the obtained PDF from $\tau$-averaging
($\bullet$) does not fit well with a modified J\"{u}ttner function
(solid green line) but fits well with a J\"{u}ttner function
divided by $\gamma$ (dashed blue line).

\smallskip
Note that the difference of temperature parameters of J\"{u}ttner
and modified J\"{u}ttner distributions in the above argument is a
result of using same expectation values in the energy constraints
in maximum relative entropy principle. It may be argued
that this assumption is incorrect due to the fact that the two
distributions refer to two different hyper surfaces in space-time
\cite{Cub09}. Undoubtedly, a suitable choice of energy expectation
value in Eq.(\ref{beta}) gives the correct temperature parameter
consistent with simulation data (i.e., $\beta=\beta_{J}$). The
important question, however, is how should one obtain the new
energy expectation value, consistent with Lorentz invariant
measure, based on fundamental laws of relativity or statistical
mechanics? As far as maximum entropy principles are considered,
the relevant constraints should be specified with respect to the
accessible information. Does changing the reference measure
introduce new information that must be adapted as a constraint?

\smallskip
At this stage we make no effort to answer these open questions and
accept (at least with regard to numerical data) that the temperature
parameter of both distribution equals $\beta_{J}$. By this choice
we still have two scenarios, first the $\tau$-averaged PDF is
described by a modified J\"{u}ttner with $\beta=\beta_{J}$ and
second, it is a rescaled J\"{u}ttner function. These two are indistinguishable
in the rest frame. However, because of different transformational properties
of J\"{u}ttner and modified J\"{u}ttner functions under Lorentz boost, they
take on distinctly different functional forms. We next consider our model in a
moving frame in order to bias this distinction.

\smallskip
\textit{Moving frame--} To this end, we examine the system for an
observer who is moving with a uniform velocity $u$ in the negative
$x$-direction with respect to the rest frame. Using the relative
entropy maximization principle, the stationary distribution will
be determined by an additional constraint on the system, namely,
that of a definite total momentum $\mathbf{P}'$ \cite{Pat72}. We
therefore obtain:

\begin{equation}\label{movfv}
    f'_{J}(\mathbf{p}')=\frac{1}{\gamma(u)Z_{J}}\exp[-\beta_{J}
\gamma(u)(E'-\mathbf{u}.\mathbf{p}')]
\end{equation}

\begin{equation}\label{movfmv}
    f'_{MJ}(\mathbf{p}')=\frac{1}{\gamma(u)Z_{MJ}}
    \frac{\exp[-\beta_{MJ}\gamma(u)(E'-\mathbf{u}.\mathbf{p}')]}
    {\gamma(u)(E'-\mathbf{u}.\mathbf{p}')}.
\end{equation}
The primed quantities are measured in the moving frame and the
additional $\gamma(u)$ term in the denominator, is due to the
contraction of the moving box that encloses the system
\cite{Kam69}. Note that the reference density in the moving frame
is obtained by $\rho'(\mathbf{p}')=\mathcal{J}\rho(\mathbf{p})$,
in which $\mathcal{J}=|\partial p^{\mu}/\partial
p'^{\nu}|=\det(a^{\mu}_{\nu})$ is the Jacobian of the coordinate
transformation, $p'^{\mu}=a^{\mu}_{\nu}p^{\nu}$. Here,
$\mathcal{J}$ equals unity since for all proper Lorentz
transformations, like boost and rotation, the determinant of the
transformation coefficients is $\det(a^{\mu}_{\nu})=1$ \cite{Wei72}.

\begin{figure}

  \includegraphics[width=8cm]{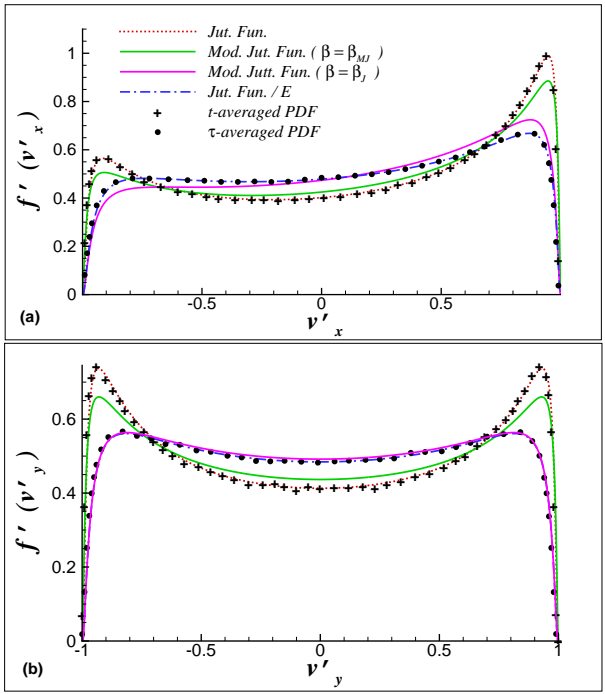}\\
  \caption{Equilibrium velocity distributions in the moving frame
  with relative velocity $u=0.1$. The system parameters
  are the same as Fig.1. In particular, $\beta_{J}=7.62$ and
  $\beta_{MJ}=4.86$. Part (a)
  shows the $x$ component and (b) shows the $y$ component distributions. The solid lines associated with
  modified J\"{u}ttner function with either $\beta_{J}$ and $\beta_{MJ}$ do not fit the data well, while J\"{u}ttner function
  divided by $E'$ with $\beta_{J}$ fits perfectly well.}
\end{figure}

\smallskip
Figure 2 shows simulation results for a system described in Fig.1
with $u=0.1$. Here, $t$-averaged PDF is obtained by measuring
velocities simultaneously with respect to the moving observer as
described in \cite{Cub07,Mon09}. Considering the Lorentz
invariance of proper-time it is not difficult to find the
$\tau$-averaged PDF in the moving frame. In this case, the
appropriate instant of measurement is the same as that of the rest
frame, however, the velocities must be recorded as seen by the
moving observer. To provide a better understanding of the effect
of motion on PDFs we have shown the $x(y)$ component of velocities
separately. The dotted (red) and solid (green and pink) lines are
the $x(y)$ component of velocity distribution obtained,
respectively, by integrating Eqs.(\ref{movfv}) and (\ref{movfmv})
over $v_{y}(v_{x})$. We have also used the fact that the
temperature parameters in the moving frame are the same as that of
the rest frame \cite{Cub07,Mon09}. As clearly seen from numerical
results, our two dimensional model shows that $t$-averaged PDF (+)
is fitted to the expected $x(y)$ component of J\"{u}ttner function
(dotted red line). The behavior of $\tau$-averaged PDF
($\bullet$), however, shows an evident deviation from the modified
J\"{u}ttner function (solid green and pink lines). This result
shows that no form of modified J\"{u}ttner function can fit the
numeric data obtained by proper time averaging. But how is a
$\tau$-averaged PDF described in the moving frame?

\smallskip
According to the arguments presented in the previous section, we
consider a J\"{u}ttner function in the moving frame divided by
$E'$, i.e.,

\begin{equation}\label{movfmmv}
    f'(\mathbf{p}')=\frac{1}{\gamma(u)Z}\frac{\exp[-\beta_{J}
\gamma(u)(E'-\mathbf{u}.\mathbf{p}')]}{E'},
\end{equation}
where $Z$ is the normalization constant. As shown in Fig.2, the
$x(y)$ component of the above distribution (dashed blue line),
fits to simulation data as expected. These results again confirm
that $\tau$-averaged PDFs are described by J\"{u}ttner functions
which are simply divided by $E$. The similarity of such functional
form to the modified J\"{u}ttner function, especially in the rest
frame, has recently led to misleading conclusions
\cite{Dun09,Cub09}.

\smallskip
\textit{Concluding remarks--} In the light of the above
discussions one might now ask what is the correspondence between
symmetries that lead to different choice of reference density in
relative entropy with the symmetries that lead to the choice of
time parameterization? In \cite{Cub09} it is argued that the choice of
coordinate time, $t$, and constant reference density results in a
J\"{u}ttner distribution while a relativistically invariant
measure along with a relativistically invariant time (i.e., proper
time $\tau$) leads to a modified J\"{u}ttner distribution. We, however, believe
that the connection between various reference densities and time
parameterization has not been well-established yet. Certainly
a MREP should directly and uniquely lead to the equilibrium
properties (i.e., PDF) once the reference density and constraints
have been specified. To the best of our knowledge, there is no proof that choice
of coordinate time leads to a constant reference density, just as no proof exists
that choice of proper time leads to $\rho=1/E$. Here, we have
shown that if one calls what results from choosing different
$\rho$'s in the MREP as J\"{u}ttner and modified J\"{u}ttner
functions, the ``modification'' that results (due to the choice of
$\rho$) in this method is different from the modification that
results from a mere reparameterization of time (which is in
accordance with Eq.(9)).

\smallskip
The authors kindly acknowledge the support of Shiraz University
Research Council.

\bibliographystyle{apsrev}
\bibliography{xbib}
\end{document}